\documentclass[lettersize,journal]{IEEEtran}
\usepackage{amsmath,amsfonts}
\usepackage{algorithmic}
\usepackage{array}
\usepackage[caption=false,font=normalsize,labelfont=sf,textfont=sf]{subfig}
\usepackage{textcomp}
\usepackage{stfloats}
\usepackage{url}
\usepackage{multirow}
\usepackage[linesnumbered,ruled]{algorithm2e}
\usepackage{verbatim}
\usepackage{graphicx}
\usepackage{cite}

\begin{document}
\bstctlcite{IEEEexample:BSTcontrol}
\title{Distributed Time Synchronization in NOMA-Assisted Ultra-Dense Networks}
\author{D. Goswami,~\IEEEmembership{Member,~IEEE}, I. Dey,~\IEEEmembership{Senior Member,~IEEE}, N. Marchetti,~\IEEEmembership{Senior Member,~IEEE}, and
S. S. Das,~\IEEEmembership{Senior Member,~IEEE}
\thanks{D. Goswami is with National Institute of Technology Calicut, India (Email: debjani.ami.89@gmail.com).} 
\thanks {I. Dey is with Walton Institute for Information and Communication Systems Science, Waterford, Ireland (Email: indrakshi.dey@waltoninstitute.ie)} 
\thanks{N. Marchetti is with Trinity College Dublin, Dublin, Ireland (Email: nicola.marchetti@tcd.ie)}
\thanks{S. S. Das is with Indian Institute of Technology Kharagpur, India (Email: suvra@gssst.iitkgp.ernet.in)}}
\maketitle
\begin{abstract}
Ultra-dense networks (UDNs) represent a transformative access architecture for upcoming sixth generation (6G) systems, poised to meet the surging demand for high data rates. Achieving precise synchronization across diverse base stations (BSs) is critical in these networks to mitigate inter-cell interference (ICI). However, traditional centralized synchronization approaches face substantial challenges in dense urban, including limited access to Global Positioning System (GPS), dependence on reliable backhaul, and high signaling overhead demands. This study advances a low-complexity distributed synchronization solution. A primary focus is on assessing the algorithm's accuracy incorporating the effects of information exchange delays, which are pronounced in large-networks. Recognizing the pivotal role of neighbor-gathered information in the proposed approach, this research employs uplink Non-Orthogonal Multiple Access (NOMA) to reduce message-gathering delays between transmitters (TXs) and receivers (RXs). 
The proposed algorithm is evaluated to assess effectiveness under exchange delays, analyzing impact of system parameters like network connectivity, size, sub-bands, etc., on synchronization speed. The findings demonstrate that the NOMA-based information-gathering technique significantly accelerates network synchronization compared to orthogonal access schemes. This advancement is crucial for meeting the low-latency requirements of beyond fifth generation (5G) systems, underscoring the potential of distributed synchronization as a cornerstone for next-generation UDN deployments.
\end{abstract}


\section{Introduction}
\begin{figure}
\centering
\includegraphics[width=0.99\columnwidth]{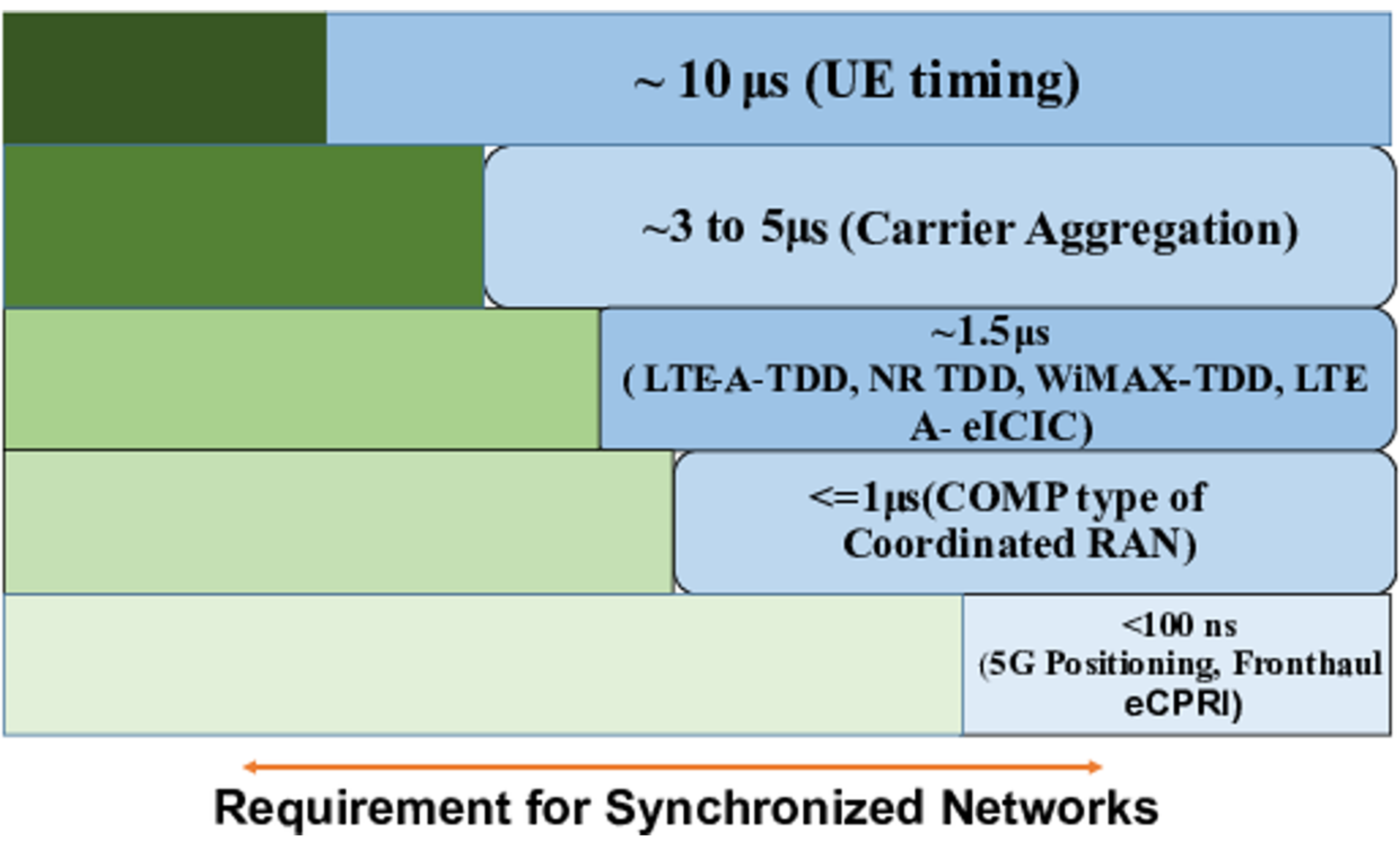}
\caption{Radio Access Network Synchronization Requirements for beyond 5G networks \cite{ITU_T_2012}.}
\label{Sync Requirement}
\end{figure}

\IEEEPARstart{T}{he} next generation of wireless communication networks, encompassing 5G, 6G and beyond, is projected to handle significantly larger traffic volumes and deliver markedly higher user data rates compared to current 4G systems. Addressing this escalating demand for data-intensive applications within the constraints of limited and costly spectrum resources necessitates the efficient utilization of all available frequency bands. Consequently, enhancing spectral efficiency (SE) has become a critical priority, prompting network architects to adopt UDN deployments featuring densely distributed small cell base stations (SBSs) alongside existing macro cells. The coexistence of small and macro cells within UDNs is facilitated by Long-Term Evolution-Advanced enhanced Inter-Cell Interference Coordination (LTE-A-eICIC). However, the reliable operation of such heterogeneous BSs hinges on precise network synchronization. Timing mismatches among BSs exacerbate ICI, significantly impairing network performance. To mitigate this heightened ICI, it is imperative to maintain a unified and accurate time scale across all BSs, underscoring the critical role of synchronization in beyond 5G networks \cite{Microsemi}. Figure \ref{Sync Requirement} highlights diverse application scenarios in beyond 5G systems that demand highly accurate timing synchronization, clearly illustrating the necessity of faster synchronization mechanisms to ensure seamless network operation.
While the Network Time Protocol (NTP) has long been regarded as a standard method for time synchronization, its susceptibility to delay attacks—caused by factors such as jamming, interference, spoofing, and signal blockage—renders it unsuitable for ultra-dense and high-reliability networks. Similarly, traditional synchronization techniques employed in LTE-Advanced systems, including ``Global Navigation Satellite Systems (GNSS) everywhere", ``Precision Time Protocol (PTP) with full path support (G.8275.1)", and ``PTP with partial on-path support (G.8275.2)", depend on Universal Time Coordinated (UTC) data from GNSS. However, in dense urban environments, where GNSS signals are frequently disrupted or inconsistent, these methods become unreliable. Furthermore, the implementation of PTP introduces additional challenges. Reliable backhaul, such as fiber or Ethernet, is required for effective operation \cite{Backhaul}, but this dependence is often impractical in urban or resource-constrained scenarios. Variations in queuing delays, packet delays, and inconsistencies across sources further degrade the timing accuracy achievable through PTP. Addressing these limitations demands proprietary adjustments and on-site calibration, which increase both operational complexity and costs. Moreover, the significant signaling overhead associated with PTPv2.1 (IEEE 1588v2) makes it ill-suited for small-cell environments \cite{PTP1}. An alternative, Sync-Ethernet technology, also escalates expenses due to the requirement for Sync-E compatibility across all devices between master and slave clocks \cite{Magazine}. These challenges underscore the necessity for innovative, low-overhead synchronization mechanisms capable of addressing the stringent requirements of beyond 5G systems, particularly in ultra-dense network deployments.

The challenges identified above motivated our earlier work in \cite{My_Paper_Time_Sync}, where we proposed a GPS-independent, distributed timing synchronization solution designed to resolve synchronization speed degradation in large networks, a limitation highlighted in \cite{Simeone2007}. Our proposed algorithm in \cite{My_Paper_Time_Sync} achieves global network synchronization by exchanging local timing information within an expanded neighborhood at minimal signaling expense. However, in our prior study, we assumed negligible time for exchanging timing information across the network. While this assumption is valid for small-scale networks, it may result in discrepancies when applied to larger networks, where the time required for information exchange can significantly affect performance. This realization prompted the present study, where we reassess the efficacy of the proposed algorithm by explicitly incorporating information exchange delays into the analysis. In \cite{My_Paper_Time_Sync}, the distributed algorithm considered both operational periods and the data transfer phase. Synchronization at each stage was achieved by utilizing local timing data collected during the previous synchronized network’s data transfer phase, in conjunction with information from the current stage. This approach underscores the critical importance of minimizing information exchange delays, as doing so not only accelerates synchronization but also extends the effective access time for system users to available resources, thereby improving network SE.

Accordingly, the primary contribution of this paper is to reduce information exchange delays in dense network scenarios and integrate this key factor into the reevaluation of the proposed algorithm’s effectiveness. To achieve this, we revisit several well-established studies that aim to minimize task completion delays within networks, identifying approaches that can enhance synchronization speed and contribute to the realization of efficient and robust UDN deployments.

\section{Related Works}
Multiple research investigations confirm that non-orthogonal multiple access (NOMA) shows significant potential for minimizing access delay. The authors in \cite{NOMA_Delay_Min_MEC_1, NOMA_Delay_Min_MEC_2} reduce task completion delays in a multi-user NOMA-Mobile Edge Computing (MEC) network. 
In particular,  the work in \cite{NOMA_Delay_Min_MEC_2} aims to minimize the task offloading delay in NOMA- assisted mobile edge computing (NOMA-MEC). The authors utilize Dinkelbach's method and Newton's method to reframe the delay minimization problem as fractional programming. They demonstrate the optimality of these approaches and investigate their rate of convergence. Notably, this study demonstrates that when sufficient energy is available for MEC offloading, pure NOMA can achieve better performance than hybrid-NOMA \cite{NOMA_Delay_Min_MEC_2}.

The authors in \cite{NOMA_Delay_Min_3} devised a combined User Pairing and Scheduling (UPaS) approach tailored for multi-carrier (MC)-NOMA systems. However, to simplify the suggested framework, the authors assume that each user experiences an identical user gain across all channels, thereby overlooking the channel-diversity advantage in orthogonal frequency-division multiplexing systems (OFDM). Considering this crucial factor into account, authors in \cite{LEO_NOMA} focus on minimizing transmission time in low earth orbit (LEO) satellite-terrestrial integrated networks. In \cite{LEO_IoT}, the authors examine strategies for minimizing the maximum data transmission completion time of Internet of Things (IoT) devices operating within combined LEO satellite-terrestrial integrated networks (STINs). In these systems, IoT devices utilize NOMA technology to send information to central earth stations (CESs), while the data transfer from CESs to the LEO satellite employs the OMA method.

Authors in \cite{Max_Completion_Time} tackles the challenge of minimizing maximum completion time in uplink multi-subcarrier NOMA networks. The authors address a complex optimization problem that combines power allocation, user pairing, and scheduling. This integrated approach is demonstrated to be nondeterministic polynomial-time (NP)-hard, highlighting the difficulty of finding optimal solutions in such networks. However, they present the optimal solution for user scheduling protocol only for cases where $K=2M$, where $K$ represents number of total devices and $M$ represents the number of subcarriers.
It is evident that the attainable data rate of an individual user in a NOMA-driven system is influenced by both its own channel condition and that of the paired user within the same resource block. Therefore, considering peer effects in the user selection procedure is crucial in this context, a factor overlooked in  \cite{NOMA_Delay_Min_3} and \cite{LEO_NOMA}. These peer effects, or externalities, add intricacy to the widely used many-to-one matching algorithm-based scheduling solution as they introduce uncertainty regarding the stability of the matching solution \cite{Peer_Effect_Imp}. Investigation in \cite{V2X_NOMA} exploits NOMA to enhance the packet reception probability. Specifically, they propose a centralized NOMA-based vehicle-to-anything (V2X) user scheduling and resource allocation technique, where each vehicle autonomously selects its transmitted power in a decentralized manner. 

The article \cite{TNSM} draws attention to the research presented in \cite{TNSM_Cited}, which examines the optimization of offloading latency and power consumption for mobile devices. This optimization aims to strike a balance between delay and energy in a mobile edge computing network enhanced by Intelligent Reflecting Surface (IRS) technology and utilizing NOMA. In \cite{TETCOM}, a joint optimization framework is developed to minimize federated learning iteration latency by managing the parameter-server’s downloading-transmission duration, devices' uploading-transmission duration as a NOMA cluster, local processing rates, and transmit powers for uploading and jamming.
Table  \ref{tab:1} presents a thorough examination of ongoing studies in this area, along with their current constraints. Inspired by the favorable results of employing NOMA to reduce task completion delay in the above study, our study exploits an uplink NOMA system for faster exchange of timing information within networks. Table 1 illustrates similarities between our study and those presented in \cite{Peer_Effect_Imp} and \cite{V2X_NOMA}. We will now highlight the distinctions between our research and these two publications. The study in \cite{Peer_Effect_Imp} aims to optimize both sub-channel allocation and power distribution simultaneously, with the goal of maximizing the weighted total sum-rate while considering user fairness. In contrast, our research focuses on minimizing task completion time in a NOMA-assisted timing synchronization solution. Although the research in \cite{V2X_NOMA} utilizes NOMA techniques to reduce latency and improve the probability of successful packet reception, it fails to address the co-channel user peer effect in NOMA-assisted systems. This effect, where scheduling one user in a sub-band (SB) impacts the achievable rates of other users in the same SB, is a critical factor that should be taken into account.
\begin{table*}[t]
  \centering
  \caption{Summarization of Related works. }
\begin{tabular}{|p{20mm}|p{35mm}|p{20mm}|p{60mm}|p{25mm}|}
 \hline
    \textbf{Research article} & \textbf{Task Offloading Minimization} & \textbf{Applications of NOMA}    & \textbf{Matching based User pairing} & \textbf{Peer Effect Consideration}\\   \hline
    \cite{NOMA_Delay_Min_MEC_1} &  ~~~~~~~~~~~~~~\checkmark & ~~~~~~~~~~~~~~\checkmark &~~~~~~~~~~~~~~$\times$ &~~~~~~~~~ $\times$ \\
    \hline
    \cite{NOMA_Delay_Min_MEC_2} & ~~~~~~~~~~~~~~\checkmark & ~~~~~~~~~~~~~~\checkmark & ~~~~~~~~~~~~~~$\times$ & ~~~~~~~~~~$\times$ \\
    \hline
    \cite{NOMA_Delay_Min_3} & ~~~~~~~~~~~~~~\checkmark & ~~~~~~~~~~~~~~\checkmark & \text{Shortest processing time algorithm based user pairing} & ~~~~~~~~~~$\times$ \\
    \hline
   \cite{LEO_NOMA} &~~~~~~~~~~~~~~\checkmark & ~~~~~~~~~~~~~~\checkmark & ~~~~~~~~~~~~~~\checkmark & ~~~~~~~~~~$\times$ \\ \hline
    \cite{LEO_IoT} &~~~~~~~~~~~~~~\checkmark & ~~~~~~~~~~~~~~$\times$ & Greedy based solution &~~~~~~~~~~$\times$ \\ \hline
     \cite{Max_Completion_Time} &~~~~~~~~~~~~~~\checkmark & ~~~~~~~~~~~~~~\checkmark & \text{Binary search based solution} & ~~~~~~~~~~$\times$ \\ \hline
     \cite{Peer_Effect_Imp} & Maximization of weighted sum rate & ~~~~~~~~~~~~~~\checkmark & ~~~~~~~~~~~~~~\checkmark & ~~~~~~~~~~\checkmark \\ \hline
      \cite{V2X_NOMA} & ~~~~~~~~~~~~~~\checkmark & ~~~~~~~~~~~~~~\checkmark & Stable room mate based matching & ~~~~~~~~~~$\times$ \\ \hline
      \cite{TNSM_Cited}  & ~~~~~~~~~~~~~~\checkmark  & ~~~~~~~~~~~~~~\checkmark &  ~~~~~~~~~~$\times$ & ~~~~~~~~~~$\times$ \\ \hline
      \cite{TETCOM}  & ~~~~~~~~~~~~~~$\times$  & ~~~~~~~~~~~~~~\checkmark &  ~~~~~~~~~~$\times$ & ~~~~~~~~~~$\times$ \\ \hline
      
  \end{tabular}
  \label{tab:1}
\end{table*}

\emph{Contribution} - In this paper, we enhance the distributed timing synchronization algorithm from \cite{My_Paper_Time_Sync} by addressing its limitations and extending its applicability to real-world, large-scale networks. While the original algorithm excelled in small-scale scenarios, its performance in complex networks, where information exchange delays are significant, necessitated further refinement. Leveraging the potential of NOMA systems to reduce task completion delays, this work optimizes synchronization efficiency by incorporating exchange delays and refining resource allocation strategies. We propose a novel framework that integrates uplink NOMA with the distributed synchronization algorithm, enabling faster and more reliable synchronization in small-cell networks. Through comprehensive analysis and numerical evaluation, we explore the effects of key parameters, including network size, connectivity, channel conditions, and subband (SB) availability, providing a robust foundation for scalable and efficient synchronization solutions in ultra-dense, beyond 5G and 6G systems. The major contributions of this paper are summarized as follows:
\begin{enumerate}
\item \textbf{Integration of Information Exchange Delay into the Synchronization Algorithm} - We extend the distributed timing synchronization algorithm proposed in \cite{My_Paper_Time_Sync} by incorporating the impact of information exchange delays, a critical consideration for real-world implementations. To minimize these delays in small-cell networks, we leverage an uplink NOMA system. Specifically, we employ the stable marriage algorithm to design an efficient association strategy between SBs and transmitting small base stations (SBSs), ensuring optimal utilization of network resources.
\item \textbf{Reevaluation of Synchronization Speed Across Various Network Configurations} - Under the proposed framework, we reassess the network synchronization speed of the algorithm in \cite{My_Paper_Time_Sync}, accounting for varying network connectivity, network size, the number of available SBs, and channel conditions. This reevaluation adapts the previous algorithm to large-scale networks, where the time required for information exchange becomes a critical factor, thereby enhancing its applicability to more complex scenarios.
\item \textbf{Performance Comparison Between NOMA and OMA Systems} - Expanding on our prior work in \cite{My_Paper_Time_Sync}, we analyze and compare the synchronization durations achieved in NOMA and OMA systems. Our findings reveal that the NOMA-based approach offers significant performance improvements in large networks, particularly in scenarios with fewer SBs, superior channel gains, and a higher prevalence of line-of-sight (LOS) pathways.
\item \textbf{Impact of System Parameters on Synchronization Performance} - We conduct an extensive investigation into how key system parameters—such as network connectivity, network size, the number of SBs, and channel conditions—affect the synchronization rate. This analysis provides a detailed discussion of observed parametric trends and their underlying causes, offering valuable insights for future system design.
\item \textbf{Stability and Computational Complexity Analysis} - We explore the stability of our proposed peer-effect-aware matching algorithm, highlighting its robustness in handling co-channel interference. We also analyze the computational complexity of the proposed solution and benchmark its efficiency against other state-of-the-art approaches, demonstrating its scalability and practicality for ultra-dense network deployments.
\end{enumerate}

The remainder of this paper is organized as follows: Section II reviews related works, highlighting the advancements and limitations of existing synchronization and resource allocation techniques in NOMA-based systems. Section III presents the system model, detailing the assumptions and configurations underpinning the proposed framework. Section IV introduces the proposed solution for minimizing information exchange delays, emphasizing the integration of uplink NOMA and stable matching algorithms. Section V provides a comprehensive numerical investigation, analyzing the impact of various system parameters on synchronization performance. Section VI discusses the stability of the proposed algorithm, while Section VII evaluates its computational complexity and contrasts it with alternative approaches. Finally, Section VIII concludes the paper, summarizing key findings of this work.

\section{System Model}
Building upon our prior research in \cite{My_Paper_Time_Sync}, this investigation considers a homogeneous small cell network scenario with $K$ small cells (picocells), denoted as $\kappa={\{1,2,3,...,K\}}$, all operating at the same transmit power, $P_t$. The S1 interface, used in LTE networks to connect the eNodeB (BS) and the core network, is used here to establish connections between the mobile core networks and the corresponding picocells. Furthermore, the X2 interface, which interconnects two eNodeBs in an LTE network, facilitates direct connections among SBSs \cite{DOMENICO20131}. In the subsequent discussion of this article, the terms "SBSs" and "nodes" will be used interchangeably. 
At each discrete time instance $n$, every SBS indexed by $k \in \kappa$ follows its local clock timing denoted by $t^{k}(n)$ where $n \in {\{0,1,2,3,...}\}$. It disseminates local timing information within its neighborhood by emitting a periodic sequence of impulses that encapsulate its timing data. 

At the nth time instant, only the incoming neighboring SBSs, denoted as $N^I_i(n)$, can decode the timing information of a specific SBS, $i$. This relationship is established as:
\begin{equation}
N^{I}_{i}(n)= \{j:P_{ij}(n)\geq P_{0}\};~i,j\in \{1,...,K\}~\text{and}~ i\neq j
\label{Incoming neighbour}
\end{equation} 
In this context, $P_{ij}(n)$ denotes the signal strength received by the $i^{th}$ SBS from the $j^{th}$ SBS at time instance $n$, while $P_{0}$, the power threshold, regulates the connectivity within the interference network.
Equation (\ref{Incoming neighbour}) suggests that the $j^{th}$ node is considered an incoming neighbor of the $i^{th}$ node only when the interfering signal from the former to the latter surpasses the established power threshold, $P_0$. In a similar manner, the set of outgoing connections for the $i^{th}$ SBS at time $n$, denoted as $N^{O}_{i}(n)$, is defined as follows:
\begin{equation}
N^{O}_{i}(n)= \{k:P_{ki}(n)\geq P_{0}\};~i,k\in \{1,...,K\}~\text{and}~i\neq k
\label{Outgoing Neighbor}
\end{equation}
We employ a Rayleigh fading channel model, taking into account the presence of non-line-of-sight (NLOS) communication in outdoor deployment. The square of the channel coefficient between the $j^{th}$ and $i^{th}$ SBS, represented by $|h_{ij}|^2$, is symbolized as $G_{ij}$ and referred to as the channel gain. This gain is characterized as a random variable following an exponential distribution.
The interference received from the $j^{th}$ SBS to the $i^{th}$ SBS, can be calculated as follows:
\vspace{0 mm}
\begin{equation}
P_{ij}(n)= P_{t}|h_{ij}(n)|^2d_{ij}^{-\alpha} = P_{t}G_{ij}(n)d_{ij}^{-\alpha}
\label{Received power}
\end{equation}
where $\alpha$ represents the path loss exponent and $d_{ij}$ is the distance between the $i^{th}$ SBS and the $j^{th}$ SBS. Consider a specific time instance $n$ with a predetermined connectivity network depicted as $G(n) =\left(\kappa,E(n),\textbf{A}(n)\right)$. In this representation, $\kappa$ nodes symbolize a set of SBSs, while $E(n)$ and $\textbf{A}(n)$ indicate the edge connections and adjacency matrix, respectively, at the $n$th time instance.
The vertex count remains constant over time, with no nodes being added or removed. However, changes in the interference graph's connectivity reflect alterations in channel coefficients at various time points. Equation (\ref{weight factor}) demonstrates that the weight value, $a_{ij}$ between two SBSs in the interference graph is proportional to the strength of the detected interference signal from one SBS to another. This choice of weight can be justified as follows: The node causing more interference to a target node requires precise timing synchronization, as timing discrepancies between them will result in a significant ICI within the network. As a result, nodes generating higher levels of interference should be given greater priority when updating local clock timing of a targetted node, which is reflected in (\ref{weight factor}).
\begin{equation}
a_{ij}(n)=\frac{P_{ij}(n)}{\sum_{j \in N_{i}^{I}} P_{ij}(n)}
\label{weight factor}
\end{equation}

In our current proposed synchronization scheme, all SBSs collaboratively achieve consistent local clock timings, addressing the limitations faced by traditional protocols like the PTP \cite{Microsemi}. Our approach, as detailed in (\ref{Proposed_Update}), leverages both incoming and outgoing neighbor information to adjust individual node timings effectively. Although the Received Signal Strength Indicator (RSSI) facilitates the measurement of signal strengths from incoming neighboring devices, the asynchronous nature of the network initially restricts data acquisition from outgoing neighbors, posing a challenge for synchronization. To address this, our algorithm's initial phase assumes complete synchronization across the network, promoting seamless data exchange. Subsequently, in the asynchronous phase, each SBS fine-tunes its local timing based on the information from outgoing neighbors obtained during the previous synchronized state and the current interference data from incoming neighbors. This strategy ensures that even if initial synchronization is absent, our algorithm progressively aligns the network using the update mechanism described in (\ref{Existing_Update}), which primarily relies on RSSI measurements from incoming neighbors. This process is defined as follows:
 \vspace{-2 mm}
	\begin{equation}
	t^{k}(n+1)=t^{k}(n)+\epsilon\sum_{j=1,~j \neq k}^{K}a_{kj}(n)[t^{j}(n)-t^{k}(n)]
	\label{Existing_Update}
	\end{equation}
	\vspace{0 mm}
    
The preceding analysis highlights that each SBS can quantify the total incoming interference from neighboring nodes, while information about outgoing neighbor interference can be gathered from the previously synchronized network. To acquire data from all outgoing neighbors, it is essential to ensure that every outgoing neighboring node falls within the incoming range of a particular node. Although this configuration is not feasible for all network types, bidirectional networks allow for this arrangement, as all outgoing neighbors is same as incoming neighbors, i.e., $\left(\text{i.e.,}~\bar{N^{O}_{i}}= \bar{N^{I}_{i}};~\forall i\right)$. In any other network configuration, receiving information from an outgoing neighbor $j$ is only feasible when 
$j \in \bar{N^{O}_{k}}\cap \bar{N^{I}_{k}}$ during the $(n-1)$ time step.
After a specific node receives information from both neighboring areas, it will adjust its local clock timing based on these two sources of data for $n$ th instant of time. This adjustment is based on two factors: previously gathered data from outbound neighbors (represented by ${a}_{jk}(n-1)$) and current interference information from inbound neighbors (denoted as $a_{kj}(n)$). More in detail, each node collects information from outgoing neighbors only once during the data transfer state of previous synchronized state to explore the timing update rule stated in (\ref{Proposed_Update}).
Specifically, our proposed timing update method in \cite{My_Paper_Time_Sync} follows the local timing update rule outlined below:
 \vspace{-1 mm}
\begin{equation}
  \begin{aligned}
  t^{k}(n+1) =  &~~t^{k}(n)+\epsilon\sum_{{j\in \kappa,~j \neq  k}} \frac{a_{kj}(n)+ {a}_{jk}(n-1)}{2}\times\\&~~~~~~~~~~~~~~~~~~~~~~\left[t^{j}(n)-t^{k}(n)\right]\\
\end{aligned}
\label{Proposed_Update}
\end{equation}
where $\epsilon$ is the step size. 
Drawing inspiration from the step size range of (0,1) and considering the value employed in \cite{Simeone2007}, we opted to set the step size at 0.9 to examine the network's optimal synchronization speed.
In this given context, if $j\notin N^{I}_{k}$, then $a_{kj}(n) = 0$, and if $j \notin \bar{N^{O}_{k}}\cap \bar{N^{I}_{k}}$ at $(n-1)$ th instant of time, then ${a}_{jk}(n-1) = 0$. In this scenario, ${a}_{jk}(n-1)$ represent the elements of the $\bar{\textbf{A}}$ matrix corresponding to the outgoing neighborhood of $k^{th}$ SBS in the preceding snapshot. It's important to highlight that the $k^{th}$ SBS can only receive information from nodes $j$ where $j \in \bar{N^{I}_{k}}$ within the $\bar{\textbf{A}}$ matrix. This implies that $j \in \bar{N^{O}_{k}}\cap \bar{N^{I}_{k}}$, and $j \subseteq \bar{N^{I}_{k}}$ at $(n-1)$ th time instant. In the case of bidirectional networks, the equality holds true for $j \subseteq \bar{N^{I}_{k}}$.
The values of $a_{kj}$ are instead collected from the present adjacency matrix (\textbf{A}). The detailed explanation of the proposed distributed time synchronization algorithm can be found in our previous work in \cite{My_Paper_Time_Sync}. 

\begin{figure*}
\centering
\includegraphics[width=1.9\columnwidth]{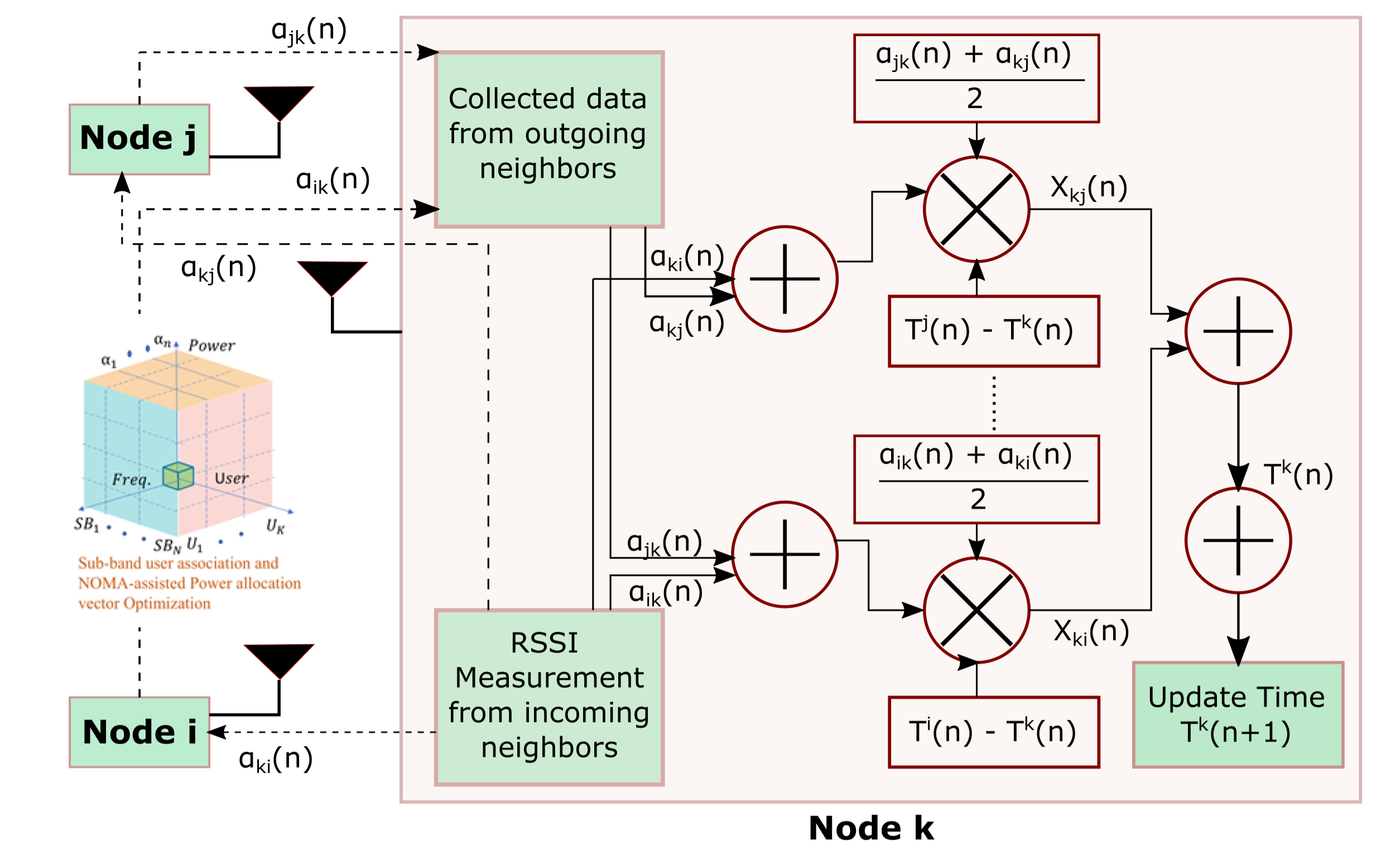}
\caption{System Model for the proposed concept.}
\label{System Model}
\end{figure*}

Our earlier study assumed that gathering information within the network took minimal time. However, it is a fact that in practical situations, collecting data across a large network is time-consuming. This temporal aspect should be factored into calculations of the actual time needed for the algorithm to synchronize the network. Consequently, this current study expands on our previous work by accounting for the time required for information exchange within the network, thus providing a more accurate representation of real-world network scenarios. The overall concept of our proposed system model in this study is illustrated in Fig. \ref{System Model}.
\section{Proposed solution to minimize information exchange Delay}
\begin{figure}
   \raggedright
\includegraphics[width=0.99\columnwidth] {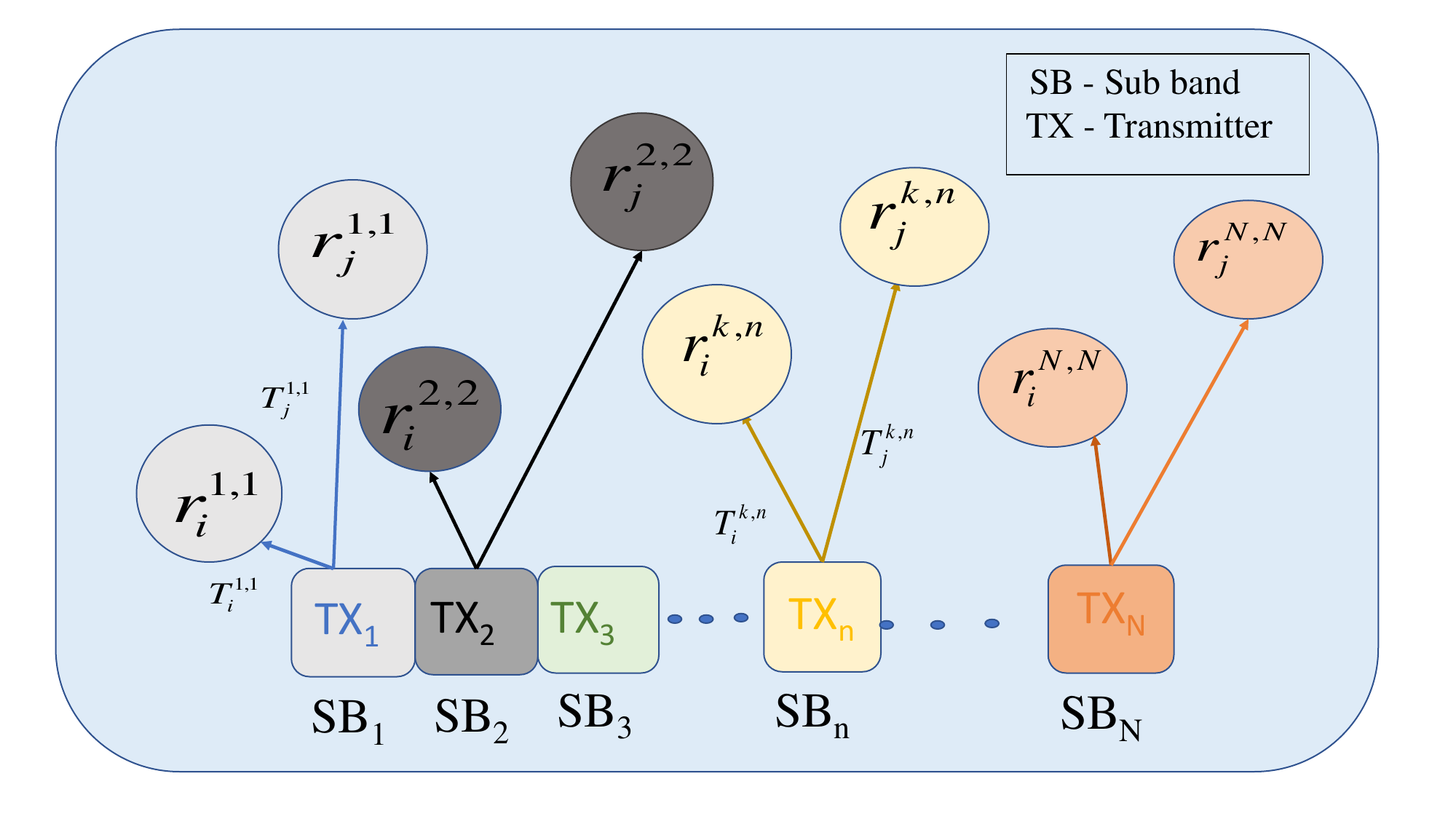}
   \caption{TX-RX Configuration}
     \label{TX-RX Config}
     \end{figure}

This work evaluates the performance of proposed synchronization algorithm while considering the objective of minimizing information exchange time with a limited number of TXs. Additionally, we assume that the system's total bandwidth, $B_\text{sys}$, is divided equally among $N$ SBs. Each SB has an individual bandwidth of $B_{sb} = \frac{B_\text{sys}}{N}$ within the frequency domain, permitting a maximum of one transmitting node to be assigned to each SB. This work adopts an uplink NOMA protocol for transmitting information from the TX to its associated RXs. Moreover, in order to restrict error propagation possibility and to minimize the Successive Interference Cancellation (SIC) decoding delay, we limit the maximum number of RXs associated with each TX to two \cite{DC-DC}. This approach can be extended to scenarios with more RXs per TX by implementing multiple time slots, where NOMA is applied to two RXs within a single time slot for each TX. Furthermore, this work considers the full-duplex mode of operation, where an individual node can transmit and receive simultaneously. 
As depicted in Fig. \ref{TX-RX Config}, this study assumes disjoint RX sets for each transmitting SBS, considering two RXs for a single TX, where each TX, $k$ scheduled on $n$th SB, sends the data towards its respective RXs $r_{i}^{k,n}$ and $r_{j}^{k,n}$ by exploiting uplink NOMA protocol. This investigation assumes that the channel gain of $r_{i}^{k,n}$ is greater than $r_{j}^{k,n}$, indicating that the $i$th node is the stronger one. It is a fact that, in the uplink NOMA network scenario, the stronger node encounters interference from the weaker node. However, the signal from the less powerful node can be decoded without interference from the comparatively stronger node \cite{Uplink_SIC_Rules}.
Hence, the superimposed transmitted signal from the $k$th TX towards its intended RX, $r_{i}^{k,n}$, scheduled on the $n$th sub-band (SB), is:
\begin{align}
y_{i}^{k,n} = f_{i}^{k,n}\left(\sqrt{\alpha_{i}^{k,n}P_{s}}S_{i}^{k,n}+ \sqrt{\alpha_{j}^{k,n}P_{s}}S_{j}^{k,n}\right) + w_{i}^{k,n} 
    \end{align}
   
Here, $f_{i}^{k,n}$ represents the complex channel response between the $k$th TX and its intended RX, $r_{i}^{k,n}$, scheduled on $n$th SB encompassing the impact of both large and small-scale fading. Additionally, $w_{i}^{k,n}$ comprises both additive Gaussian white noise and ICI. 
Furthermore, $S_{i}^{k,n}$ and $S_{j}^{k,n}$ represent complex Quadrature Amplitude Modulation (QAM) symbols drawn from a constellation with zero mean and unit variance. $P_{s}$ is the total transmit power, while $\alpha_{i}^{k,n}$ and $\alpha_{j}^{k,n}$ are the power allocation coefficients of the $k$th TX for its intended RXs $r_{i}^{k,n}$ and $r_{j}^{k,n}$, respectively, satisfying $\alpha_{i}^{k,n} + \alpha_{j}^{k,n} = 1$.
Hence, the post-SIC Signal-to-Interference plus Noise Ratio (SINR) of the strong RX, $r_i^{k,n}$, of TX $k$ scheduled in the $n$th SB is given by:

 \begin{align}
  \gamma_{i}^{k,n} = \frac{|f_{i}^{k,n}|^2\alpha_{i}^{k,n}P_{s}}{\underbrace{|f_{i}^{k,n}|^2\alpha_{j}^{k,n}P_{s}}_{\text{Co-RX interference from $r_j^{k,n}$}}~+~\sigma_{k,n}^2}
  \end{align}
Here, $\sigma_{k,n}^2 = \mathbb{E}|w_{i}^{k,n}|^2$.
Therefore, the achievable data rate for the $i$th RX of the $k$th TX allocated to the $n$th SB can be expressed as follows:
 \begin{align}
R_{i}^{k,n} = \log_2(1+ \gamma_{i}^{k,n})
\label{NOMA rate}
    \end{align}
   Similarly, the post-SIC SINR and the achievable data rate of weak RX, $r_j^{k,n}$, in the $n$ th SB is,
    \begin{align}
  \gamma_{j}^{k,n} = \frac{|f_{j}^{k,n}|^2\alpha_{j}^{k,n}P_{s}}{\sigma_{k,n}^2}\\
  R_{j}^{k,n} = \log_2(1+ \gamma_{j}^{k,n})
  \end{align}
  Note that if the $k$th TX completes its transmission to the $i$th RX, ${r_{i}^{k,n}}$ before the transmission time taken to the $j$th RX, then the $i$th RX will attain the NOMA rate, denoted as ${R_{i}^{k,n}}$, for the entire duration of its transmission. This is because the same SB ($n$th in this case) is shared between both RXs of $k$th TX for the entire transmission duration of the ${r_{i}^{k,n}}$. 
Conversely, if the transmission time to the ${r_{j}^{k,n}}$ RX is shorter than that of the $i$th RX, once the $k$th TX finishes its transmission towards $j$th RX (say, $T_{j}^{k,n}$), the transmission between $k$th TX and $i$th RX gains access to the entire $n$th SB. This allows  ${r_{i}^{k,n}}$ to achieve the data rate of OMA beyond the time of $T_{j}^{k,n}$, as indicated by Eq. (\ref{NOMA_Rate}). 
    
\begin{align}
T_{i}^{k,n} = \begin{cases} & \frac{L}{R_{i}^{k,n}}~~~~~~~~~~~~~~~~~~~~;\text{if}~~ \frac{L}{R_{i}^{k,n}}\leq T_{j}^{k,n} \\
&T_{j}^{k,n} + \frac{L-R_{i}^{k,n}T_{j}^{k,n}}{R_i^{\text{OMA},k,n}}~~~;\text{else}
\end{cases} 
\label{NOMA_Rate}
\end{align}
where, 
\begin{align}
\gamma_{i}^{\text{OMA},k,n} = \frac{|f_{i}^{k,n}|^2\alpha_{i}^{k,n}P_{s}}{\sigma_{k,n}^2}\\
  R_i^{\text{OMA},k,n}= \log_2(1+\gamma_{i}^{\text{OMA},k,n})
  \end{align}
In this scenario, $L$ represents the quantity of bits transmitted by each sender. This work assumes that all SBS transmit an identical number of bits, which is typical for Narrowband Internet of Things (NB-IoT) devices. However, it is straightforward to incorporate variable-length bits into this analysis. Hence, the overall uplink transmission time from the $k$th TX corresponding to its associated RXs  ${r_{i}^{k,n}}$ and  ${r_{j}^{k,n}}$ scheduled in the $n$th SB at time instant, $t$ is given by:
\begin{align}
T^{n}(t) = \text{max}{\{T_i^{k,n}(t),T_j^{k,n}(t)}\}
\end{align}
The maximum task completion time across all SBs at a given time instant, $t$ is given by:
\begin{align}
T_{\text{max}}(t) = \max_{n \in N} T^{n}(t)
\end{align}
This research focuses on reducing the information transmission time in each individual time slot. To achieve this goal, efficient SBS-SB scheduling and power allocation strategies have been investigated. Therefore, the objective function is defined as:
\begin{subequations}
\begin{align}
\min_{\varrho, \alpha_{k}} \quad & T_{\max}(\varrho, \alpha_{k})\label{SE_0}\\
\textrm{subject to} \quad & 
 \sum_{k \in \mathcal{K}}
  \rho_{k,n} \leq 1;~~ n \in \{1,2,..,N\} \label{SE_1}\\
  &\sum_{n \in \mathcal{N}}
  \rho_{k,n} \leq 1;~~ k \in \{1,2,..,K\}   \label{SE_2}\\
& \rho_{k,n} \in \{0,1\}\label{SE_3}\\
 &\rho_{k,n}=  \rho_{n,k}\label{SE_4}\\
& \alpha_{k,i} \geq 0\label{SE_5}\\
&\alpha_{k,j} \geq 0\label{SE_6}\\
 &\alpha_{k,i} + \alpha_{k,j} \leq  1\label{SE_7}
\end{align}
\end{subequations}
where, $\varrho$ denotes the binary assignment matrix encompassing all $K$ TXs and $N$ SBs blocks (SBs). Additionally, $\alpha_{k}$ represents a vector containing the power allocation coefficients for the intended RXs of the $k$th TX, i.e., $r_{i}^{k,n}$ and $r_{j}^{k,n}$.
In particular, $\rho_{k,n} = 1$ indicates that the $k$th TX is assigned to the $n$th SB, and it is zero otherwise. Eq. (\ref{SE_1}) ensures that each SB can be occupied by at most one TX, while Eq. (\ref{SE_2}) confirms that one TX can be allocated to at most one SB. 
Furthermore, $\alpha_{k,i}$ and $\alpha_{k,j}$ represent the power allocation coefficients for uplink NOMA protocol of the $k$th TX for its intended RXs $r_{i}^{k,n}$ and $r_{j}^{k,n}$, respectively. Equations (\ref{SE_5})-(\ref{SE_7}) confirm that the power allocation coefficients are non-negative, and their sum is constrained to unity. One can easily understand that objective function (\ref{SE_0}) is a combinatorial optimization that incorporates both discrete constraints (SBS-SB scheduling) and continuous variable constraints (power allocation). 
Therefore, to address this framework, we decouple the optimization variables and analyze them separately. While decoupling the optimization problem may sacrifice some degree of optimality, it can yield a sub-optimal solution that is more tractable and less complex than the exponentially complex optimal solution. Ultimately, the proposed solution is achieved through an iterative process.
\subsection{Small cell base station - Subband (SBS-SB) association Strategy}
\begin{figure}[t]
\centering
\includegraphics[width=0.99\columnwidth]{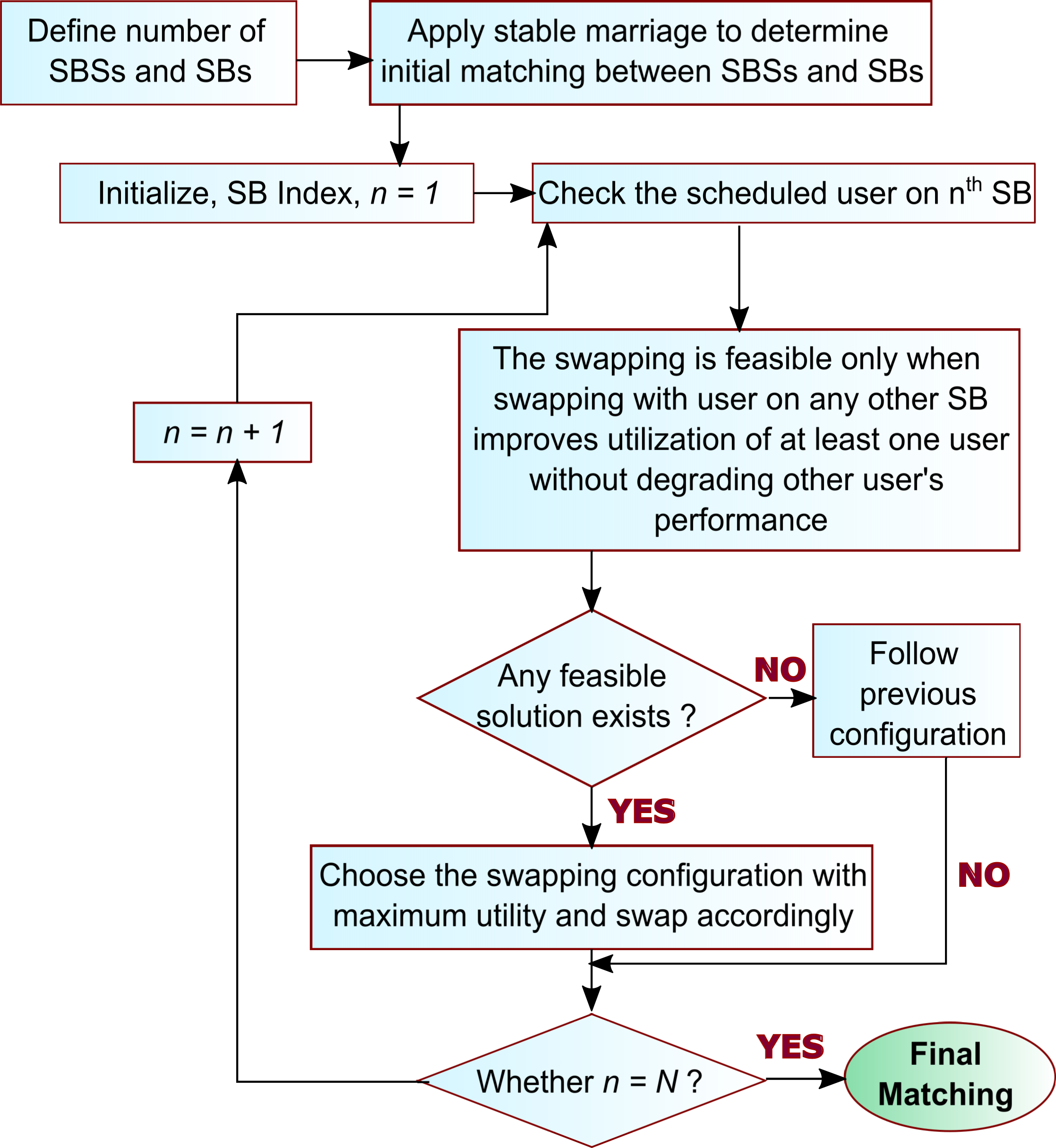}
     \caption{Flow chart of the proposed algorithm}
     \label{Flow chart}
     \end{figure}

We begin our SBS-SB association process by assuming that the power allocated for transmission from a single TX to its intended RXs remains constant, denoted as, i.e., $\alpha_{k} = \alpha_{k}^* ; ~\forall k \in \{1,2,...,K\}$. In particular, for a given power allocation strategy, say $\alpha_{k}^*$, we can reformulate the sub-problem of SB assignment as follows: 
\begin{subequations}
\begin{align}
&~~~~~~~~\min_{\varrho} \quad  T_{\max}(\varrho, \alpha_{k}^*)\\
&\textrm{subject to}~~~~~ (\ref{SE_1})-(\ref{SE_4})
\end{align}
\end{subequations}
The formulated problem exhibits non-convex characteristics, primarily attributed to the presence of interference terms in the achievable rate expression \cite{Nemhauser1988IntegerAC}.  

In this paper, we discuss the impracticality of employing an exhaustive search for optimal solutions in dense network environments due to its exponential growth in computational complexity as the number of users increases. To circumvent these challenges, we introduce a dynamic two-sided one-to-one matching process, which considers SBSs and SBs as distinct sets of players. This method is structured around the stable marriage problem, a well-suited algorithmic framework for establishing robust associations between SBSs and SBs. The stable marriage problem ensures a stable matching between SBSs and resources (SBs), effectively addressing network constraints by scheduling no more than one TX per SB, as formalized in Eqs. (\ref{SE_1}) and (\ref{SE_2}).

Moreover, our approach incorporates the unique dynamics of uplink NOMA, which involves two RXs for each TX. This setup necessitates the inclusion of the peer effect in our algorithmic design, a phenomenon where the pairing of one RX with a TX can significantly influence the achievable data rate for another RX linked to the same TX, thereby effecting the message transfer time, as these two factors are inversely related to each other. To manage this complexity, our SBS-SB scheduling procedure explicitly accounts for the peer effect, enhancing the algorithm's efficacy and stability. By integrating these considerations, our methodology not only optimizes resource allocation but also ensures that the system's performance remains robust under the varying demands of dense network configurations. This approach allows for a more nuanced understanding of network behavior, leading to more efficient and reliable system performance.
\subsection{Power Allocation Strategy}
In the next stage, given a specific SBS-SB scheduling, say, ${\varrho}^*$, we have optimized the power allocation strategy. Therefore, for a given SBS-SB scheduling policy, the sub-problem of power allocation is reformulated as follows:
\begin{subequations}
\begin{align}
&~~~~~~\min_{\alpha_{k}} \quad  T_{\max}(\varrho^*, \alpha_{k})\\
&\textrm{subject to}~~(\ref{SE_5})-(\ref{SE_7})
\end{align}
\end{subequations}
In our study, we meticulously adjust the power allocation coefficients, denoted as \(\alpha_{k,i}\) and \(\alpha_{k,j}\), within the established SBS-SB scheduling matrix, \(\varrho^*\). These coefficients are fine-tuned in minuscule increments to determine the optimal settings that minimize message transfer delay. It is crucial to recognize that the effectiveness of these coefficients is intrinsically linked to the configuration of the SBS-SB matrix, and their optimal values can vary significantly across different scheduling schemes. This dependency introduces a challenge in achieving a universally optimal power allocation across varying network conditions. To overcome this challenge, our approach involves an iterative process that refines both the SBS-SB scheduling and the power allocation iteratively. This method allows for continuous adjustments based on the dynamic network requirements, ensuring that the system adapts to achieve minimal message transfer delays efficiently. The iterative process is designed to halt when further adjustments cease to produce significant improvements in the objective function, indicating that a near-optimal configuration has been achieved.

For a detailed exposition of the methodology employed, Algorithms \ref{algo_1} and \ref{algo_2} outline the steps involved in the SBS-SB scheduling and power allocation processes. These algorithms also detail the procedures for calculating the operational duration of our algorithmic approach, ensuring a comprehensive understanding of both the implementation and the performance evaluation phases. This structured approach not only enhances the precision of our system configuration but also contributes significantly to the robustness and reliability of network operations in densely populated environments.

\begin{algorithm}[ht]
		\caption{Algorithm for SBS-SB scheduling and power allocation.}
		\label{algo_1}
		\textbf{Input:} Number of Nodes $K$, Distance matrix $(\textbf{D})_{K\times K}$, Initial time vector $(\textbf{T}^{(0)})_{1\times K}$,  Power Threshold Value, Maximum number of iterations, $N_{max}$, NOMA Power coefficient of strong RX ($\alpha_s$).\\
		\textbf{Initialize:} 
		Number of iterations, $N_{Ite}~=~0$\\
         Use Algorithm 2 to update the local timing information for all SBSs until convergence is achieved.\\
          Compute the average standard deviation across all snapshots and evaluate the algorithmic operational time.\\
         Decide the initial matching between the SBSs and the SBs using the stable marriage algorithm.\\
         \For{ $\alpha_s$ = 0:0.0025:1}
         {
          \For{$N_{Ite}$ $\leq$ $N_{max}$}
          {
          Employ a swapping strategy to identify a Pareto-optimal solution of matching algorithm, ensuring that the utility of at least one SBS improves without reducing the utility of the paired SBS scheduled in the same SB for all SBs.\\
          Select the swapping configuration with minimum information exchange delay.\\
          }
          Calculate the average time required for both NOMA and OMA-assisted message transfer schemes across all iterations for a fixed power allocation coefficient.
         
         }
      Select the power allocation coefficient that provides the minimum access delay. Record the value as the shortest possible information exchange delay.
\\
To determine the total network synchronization time, combine the algorithmic operational time derived from Algorithm 2 with the shortest possible information exchange delay. \\
         \textbf{Return:} Maximum Synchronization time for NOMA and OMA $( T_{\text{sync}}^{\text{NOMA}}, T_{\text{sync}}^{\text{OMA}})$, respectively.
	  \end{algorithm} 

 \begin{algorithm}[ht]
		\caption{Algorithm to compute the algorithmic operational time.}
		\label{algo_2}
		\textbf{Input:} Initial power matrix $(\textbf{P}^{1})_{K\times K}$, Maximum number of snapshots $(T_{max})$, Maximum number of iterations per snapshot $(n_{max}-1)$, Initial adjacency matrix $(\textbf{A}^{1})_{K\times K}$, Distance matrix $(\textbf{D})_{K\times K}$,  Maximum standard deviation (SD) allowed (${\delta}$), initial standard deviation among clock timings $C^{(0)}(\textbf{A}_{n}^{T_{Index}}, \bar{\textbf{A})}$, Initial time vector $(\textbf{T}^{(0)})_{1\times K}$, No. of nodes, Power Threshold Value.\\
		\textbf{Initialize:} 
		$\textbf{{P}}^{1}$,~$n~=~0$,~~$T_{Index}~$~=~2,	$\bar{\textbf{A}}=~\textbf{A}^{1}$,~~$C^{(0)}(\textbf{A}_{n}^{T_{Index}},  \bar{\textbf{A})}=0$,
	\\
		\For{$T_{Index}$ $\leq$ $T_{max}$}
		{
			Generation of power matrix $\left(\boldsymbol{P}^{T_{Index}}_{n}\right)$ for all SBSs by following (\ref{Received power}).\\
			Generate the adjacency matrix $\textbf{A}^{T_{Index}}_{n}$ using (\ref{weight factor}) and  the time vector  $(\textbf{T}^{(n+1)})$ by following (\ref{Proposed_Update})\\ 
			Calculate,   $C^{(n+1)}(\textbf{A}_{n}^{T_{Index}}, \bar{\textbf{A})}$ by using (\ref{SD}).\\
			\eIf{$( C^{(n+1)}(\textbf{A}_{n}^{T_{Index}}, \bar{\textbf{A}})) > \delta$ \text{and} $n < n_{max}$}
			{
				$n=n+1$;\\
				Generation of power matrix $\left(\boldsymbol{P}^{T_{Index}}_{n}\right)$ for all SBSs by following (\ref{Received power}) and corresponding adjacency matrix $\textbf{A}^{T_{Index}}_{n}$.\\
				Update the time values of all SBSs $(\textbf{T}^{(n+1)})$ using $(\textbf{T}^{(n)})$ by following 	(\ref{Proposed_Update}) \\ 
				Go to the line 6\\
			}
			{
			}
			Record $C^{(n+1)}(\textbf{A}_{n}^{T_{Index}}, \bar{\textbf{A}})$, $n$, and $\textbf{T}^{(n+1)}$  for this snapshot, $T_{Index}$.\\
				$\bar{\textbf{A}}=\textbf{A}_n^{T_{Index}}$\\
				$T_{Index}~$~=~$T_{Index}+1~$\\
			
		}
		Average the $C^{(n+1)}(\textbf{A}_{n}^{T_{Index}}, \bar{\textbf{A})}$ and $n$ over all snapshots, $T_{max}$.\\
		\textbf{Return:} SD between time values $( C_{\text{avg}}^{(n+1)}(\textbf{A}^{T_{Index}}, \bar{\textbf{A}}))$, Number of required iterations to converge  $n_{\text{avg}}$, Algorithmic operational time.
	  \end{algorithm}  
In our analysis, the distance matrix plays a crucial role as it quantifies the spatial distances between each pair of SBSs within the network. This matrix is foundational for evaluating the interaction dynamics and connectivity among the SBSs. Moreover, we introduce the concept of the Connectivity Factor (CF), which is mathematically defined as follows:

\begin{equation}
	C_F = \frac{\sum_{i=1}^{K} (|N^{I}_{i}| + |N^{O}_{i}|)}{2 {K \choose 2}}
	\label{CF}
\end{equation}

Here, \(|N^{I}_{i}|\) and \(|N^{O}_{i}|\) represent the number of incoming and outgoing connections, respectively, for the \(i\)-th SBS. The denominator, \(2 {K \choose 2}\), normalizes this sum by the total number of possible connections between pairs of stations, providing a scaled measure of overall network connectivity. This metric is instrumental in understanding the density and interconnectivity of the network, which directly influences communication efficiency and system robustness. CF also provides a quantitative measure of how well-connected the network is, which is crucial for algorithms that rely on the rapid and reliable exchange of timing information across SBSs. A higher CF indicates a denser network with multiple pathways for communication, facilitating faster dissemination and more accurate synchronization of timing data across the network.

For our proposed algorithm, which utilizes information from both incoming and outgoing neighbors to adjust the timing at each node, a higher CF means that more nodes are within direct communication range of each other, allowing for more robust and quicker synchronization. This is especially critical in environments where timing accuracy is paramount to network performance, such as in communication networks supporting real-time applications and services. By ensuring a high CF, our proposed algorithm can operate more efficiently, achieving faster convergence to a synchronized state and maintaining stability even in dynamic conditions. This ultimately enhances the overall performance and reliability of the network, demonstrating the critical role of CF in supporting sophisticated network synchronization algorithms.

Furthermore, the variability in synchronization across the network at any given instant \(t\) is captured by the standard deviation of the timing vector, which is defined by the equation:

\begin{equation}
	C^{(t)}(\textbf{A}, \bar{\textbf{A}}) = S_D(t) = \sqrt{\frac{\sum_{i=1}^{K} (t^{i}(t) - \bar{t}(t))^2}{K-1}}
	\label{SD}
\end{equation}

In this equation, \(t^{i}(t)\) denotes the timing at the \(i\)-th SBS at time \(t\), and \(\bar{t}(t)\) is the average timing across all SBSs at the same instant. This standard deviation serves as a critical measure of temporal dispersion within the network, highlighting the synchronization disparities among the SBSs. Monitoring this metric allows for targeted adjustments in network synchronization protocols, enhancing overall temporal alignment and reducing the likelihood of communication errors due to timing discrepancies. The precise understanding of these metrics facilitates the optimization of network operations and underscores the intricate dynamics of densely deployed SBS environments.

\section{Numerical Investigations}

This section examines the impact of system parameters on the effectiveness of our proposed time synchronization algorithm.
The system parameter values used in this study are presented in Table \ref{Table_SimulationParameters}.
	\begin {table}
	    \begin{center}
		\caption {Simulation Parameters.}
		\label{Table_SimulationParameters}
		\begin{tabular}{|p{25mm}|p{20mm}|p{20mm}|p{8mm}|}
			\hline
			\textbf{Parameter} & \textbf{Value} & \textbf{Parameter} & \textbf{Value}\\
			\hline
			\hline
			Number of SBSs & 250 &Path loss exponent & 4\\
			\hline
			Power Thresold & -110 dBm \cite{3gpp_36.133} &$\epsilon$& 0.9\\
			\hline
			Base station transmit power & 23 dBm \cite{3gpp_07_2018} & Node Temperature & $0-50^{\circ}$\\
			\hline
			Temperature Coefficient ($\beta$) &-.042 ppm$/^{\circ} C^2$\cite{TCXO} &Maximum SD allowed  ($\delta$) & 1e-06\\	\hline
   	 Max no. of TTI ($T_{max}$) & 1,000 & Max no. of iterations 
        ($n_{max}$) & 20,000\\
        \hline
		\end{tabular}
	\end{center}
\end{table}

This section presents our numerical investigation findings. We will initiate the analysis of our proposed algorithmic solution. Subsequently, we will explore the configuration for NOMA-based data transmission within the networks. The study examines 500 periods of synchronization, with each period allowing for up to 20,000 iterations. The synchronization algorithm (Algorithm 2) employs these iterations to minimize timing discrepancies between nodes until it either reaches convergence or hits the maximum iteration limit. To optimize the benefits provided by the NOMA system, we implement a two-tier SBS structure, consisting of near SBSs and far SBSs. Specifically, the near SBSs are positioned uniformly within a range of 0 to 10 meters, while the far SBSs are situated between 10 to 100 meters. Furthermore, we presume that the initial time discrepancies of the SBSs are evenly spread between 0 and 40$\mu$s. For further details, readers are referred to \cite{My_Paper_Time_Sync}. We employ the stable marriage algorithm to establish the initial matching between SBs and SBSs, defining the utility of SBSs by the minimum message transfer delay in the network.

In later stage, we permit configuration swapping between SBSs and SBs only if it maximizes the  utility of at least one TX-RX configuration without degrading the utility of the other RX associated with the same TX.      
\begin{figure}
\centering
\includegraphics[width=1.1\linewidth]{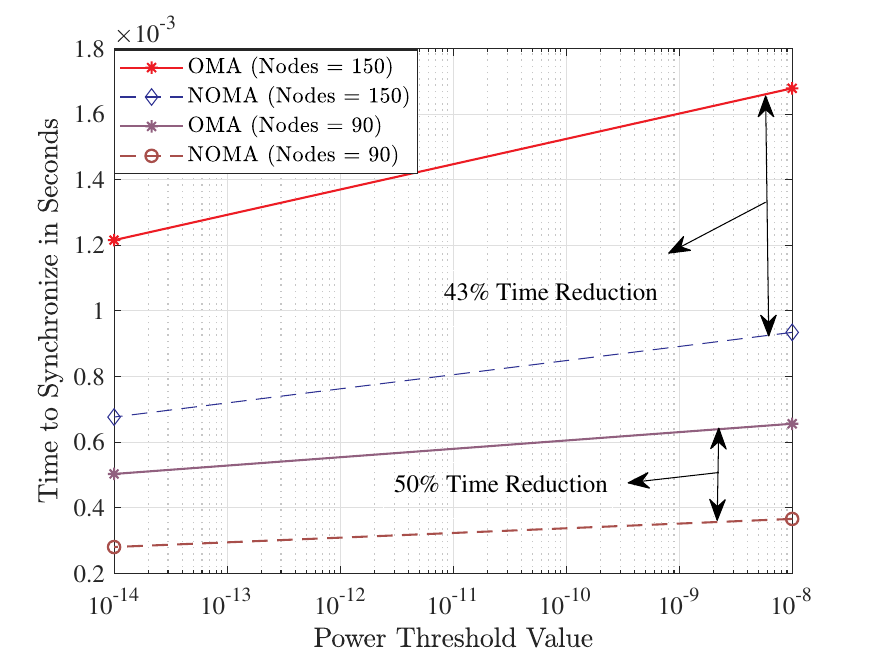}
\caption{Network synchronization time for various network connectivities controlled by power threshold value for different network sizes.}
\label{Time_Sync_Different_CFs}
\end{figure}
Fig. {\ref{Time_Sync_Different_CFs}} demonstrates the network synchronization time for various network connectivities controlled by the power threshold value, \(P_0\). Eqs. (\ref{Incoming neighbour})-(\ref{Outgoing Neighbor}) demonstrate how an increase in power threshold value reduces network connectivity. Fig. {\ref{Time_Sync_Different_CFs}} confirms that increasing the power threshold value, or, in other words, reducing network connectivity, results in an increase in the network synchronization time. This is because less network connectivity, or a more constrained neighborhood, slows down the speed of information spreading, leading to increased time required for exchanging local timing information within the network.

As mentioned earlier, our proposed algorithm divides the network synchronization time into two phases, namely the algorithmic operational phase and the information exchange phase. The network achieves synchronization at a particular stage by utilizing local timing information collected from both the current stage and the previous synchronized network's data transfer phase, as outlined in the introduction. To determine the network synchronization time, it is necessary to account for both the algorithmic operational phase and the NOMA-assisted data transfer duration from the previously synchronized network. While the algorithmic operational phase remains constant in terms of time required, the key distinction lies in the information exchange times contributed by the NOMA and OMA approaches. The reason for this is that, in contrast to the algorithmic operational time, the transmission of messages within the network heavily relies on how user-resource blocks are associated. This association strategy differs between NOMA and OMA schemes. More in detail, in this set up, the up-link NOMA scheme permits parallel transmission from one TX to two RXs within a single resource block, whereas the OMA scheme necessitates serial transmission for the same task. Consequently, it is intuitive that the NOMA access scheme will outperform the OMA scheme, or at least, achieve equivalent performance, which is reflected in Fig. {\ref{Time_Sync_Different_CFs}}. 

\begin{figure}[t]
\centering
\includegraphics[width=1.05\linewidth]{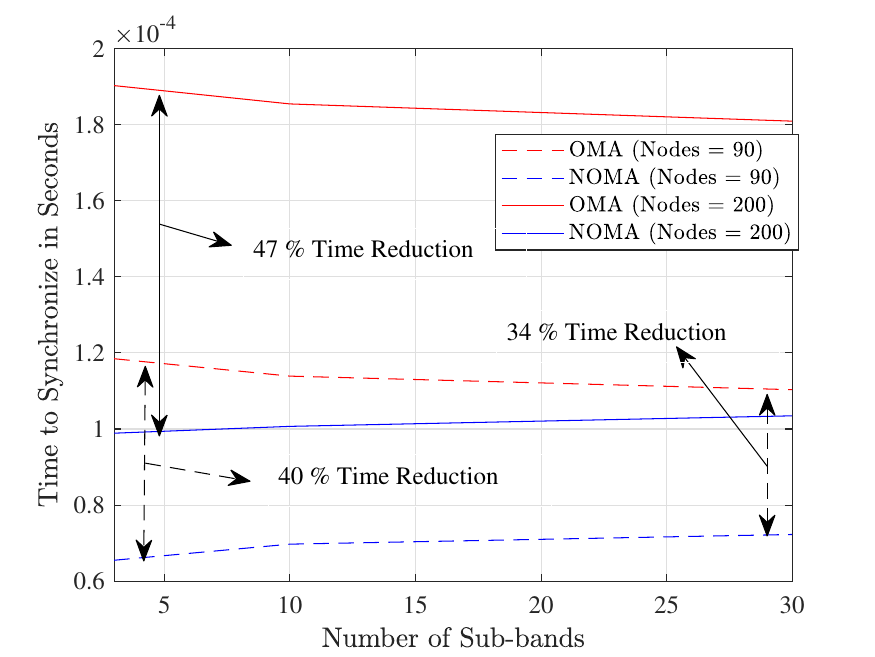}
\caption{Network synchronization time as a function of number of SBs in the system.}
\label{Time_Sync_Different_Nodes}
\end{figure}
~Fig. \ref{Time_Sync_Different_Nodes} illustrates how the number of SBs affects information spreading time within networks. Moreover, the figure emphasizes the impact of network size on information exchange delay. A network consisting of 90 nodes demonstrates a 40\% improvement in NOMA performance compared to the OMA method. This advantage increases to 47\% when the network expands to 200 nodes. Furthermore, with a fixed number of system nodes ($K$ = 90), we achieve a 40\% NOMA gain with a limited number of SBs. However, this benefit decreases to 34\% when the system's available resource blocks are increased to 30.
These findings shed light on the potentiality of NOMA in minimizing the information exchange delay, especially in scenarios with a limited number of SBs in conjunction with a large number of nodes. In contrast, the NOMA-gain diminishes as the number of available SBs in the system increases.

This phenomenon can be explained as follows: when the quantity of SBs is similar to the number of nodes in the system, an OMA system can serve the majority of nodes simultaneously by employing multiple orthogonal frequency blocks. This simultaneous transmission from limited nodes effectively reduces the need for multiple nodes to be scheduled in a single resource block, which aligns with the NOMA principle. Conversely, when the node count significantly exceeds the available system resources, multi-node scheduling within a single RB can be advantageous as it provides the opportunity for parallel transmission. Due to this reason NOMA scheme outperforms the serial transmission characteristic associated with the OMA scheme. 
\begin{figure}[t]
\centering
\includegraphics[width=1\linewidth]{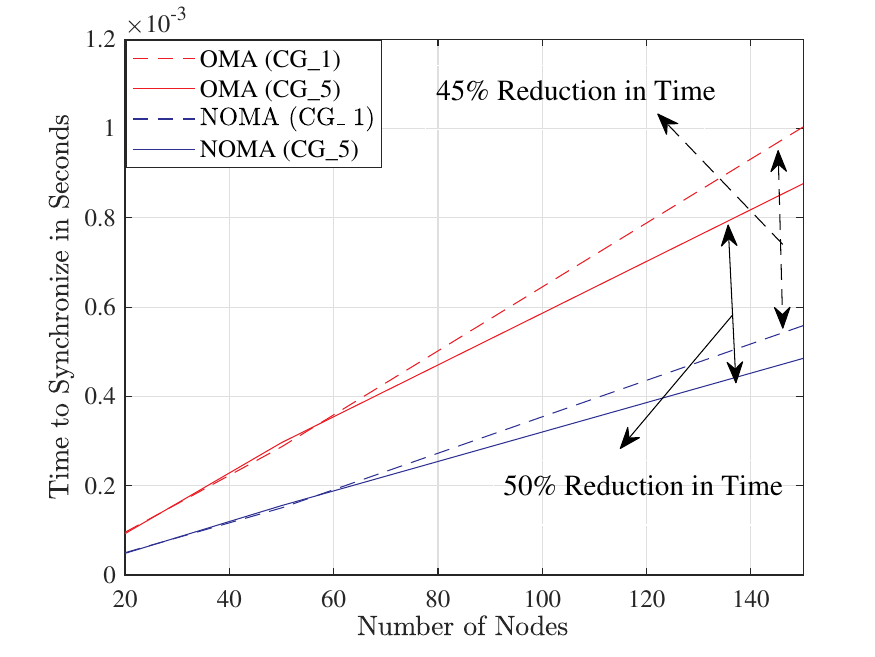}
\caption{Network synchronization time for various channel conditions.}
\label{Time_Sync_Different_CGs}
\end{figure}

Fig. \ref{Time_Sync_Different_CGs} illustrates the information exchange delay under
different channel conditions. This work considers the Rayleigh
fading channel to address NLOS scenarios in outdoor deployments, modeling the channel gain as an exponentially distributed random variable. This figure shows that in the presence of a comparatively weaker Rayleigh-faded channel,
where the channel gain has a mean of 1 in the exponential
distribution, the time required to gather information is significantly higher compared to a system with better channel
conditions (with a higher mean value of the exponential
distribution). As stated in Eq. (\ref{Received power}), a higher channel gain indicates greater network connectivity. As shown in the Fig. \ref{Time_Sync_Different_CFs}, increased network connectivity accelerates the dissemination of timing information across the network, thereby enhancing the synchronization speed. Therefore, the observed parametric trend aligns with expectations. Furthermore, it has been observed that in configurations with comparatively weaker channel conditions, the advantage of NOMA is less compared to scenarios with better channel conditions. In particular, while in strong channel condition-associated cases, we achieve a NOMA gain of 50\%, this gain decreases to 45\% for weaker channel-associated systems.
\begin{figure}[t]
\centering
\includegraphics[width=1\linewidth]{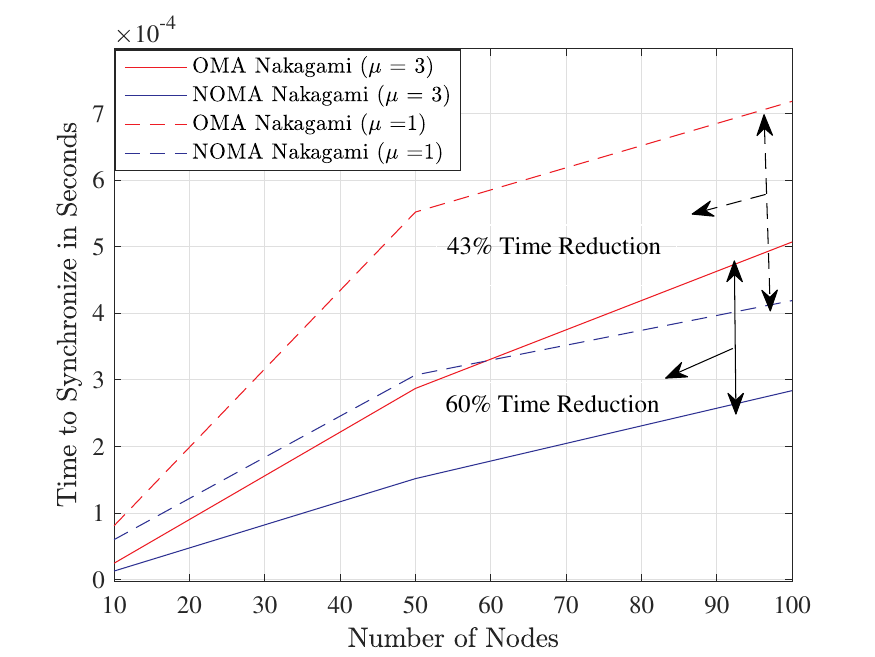}
\caption{Comparison of network synchronization time for line-of-sight (LOS) and the non-line-of-sight (NLOS) scenarios.}
\label{Time_Sync_Nakagami_ Different_Mus}
\end{figure}

 To broaden our investigation, this work examines the efficiency of our proposed solution for both LOS and NLOS scenarios in Fig. \ref{Time_Sync_Nakagami_ Different_Mus}. In particular, we modeled the LOS channel using a Nakagami distribution with a shape parameter of 3, 
and compared it to a more NLOS scenario with a shape parameter of 1. Note that a shape parameter of 1 denotes an NLOS channel, whereas higher shape parameter value ensures the presence of more LOS paths in the system. Fig. \ref{Time_Sync_Nakagami_ Different_Mus} shows that our proposed solution achieves earlier synchronization within the network when there are more LOS components (with a high shape parameter value, $m = 3$) compared to comparatively NLOS scenarios (with $m = 1$). 
An explanation for this finding can be found in \cite{Yang2017CapacityON}, which demonstrates how capacity fluctuates with an increase in the Nakagami-$m$ fading shape parameter, $m$, when the signal-to-noise ratio (SNR) is 0 dB.The authors in \cite{Yang2017CapacityON} confirms that as the $m$ value increases, moving towards the LOS scenario, the data rate also increases. Note that access delay is inversely proportional to data rate, as access delay can be defined as $\frac{L}{R}$, In this context, $L$ signifies the quantity of data bits transmitted, while $R$ represents the data rate that can be achieved. Therefore, an increase in data rate with an increasing shape parameter value automatically reduces message transfer time, thereby increasing synchronization speed in comparatively more LOS scenario. Moreover, this figure shows a significant NOMA gain of 60\% for the LOS scenario, which is reduced to 43\% for the NLOS scenario. This observation exhibits that the NOMA gain is more prominent than the conventional OMA scheme for Rician fading channels compared to Rayleigh fading channels, directly aligning with the findings of \cite{NOMA_OMA_Nakagami}. 
\begin{figure}
\includegraphics[width=8.5cm,height=6.5cm]{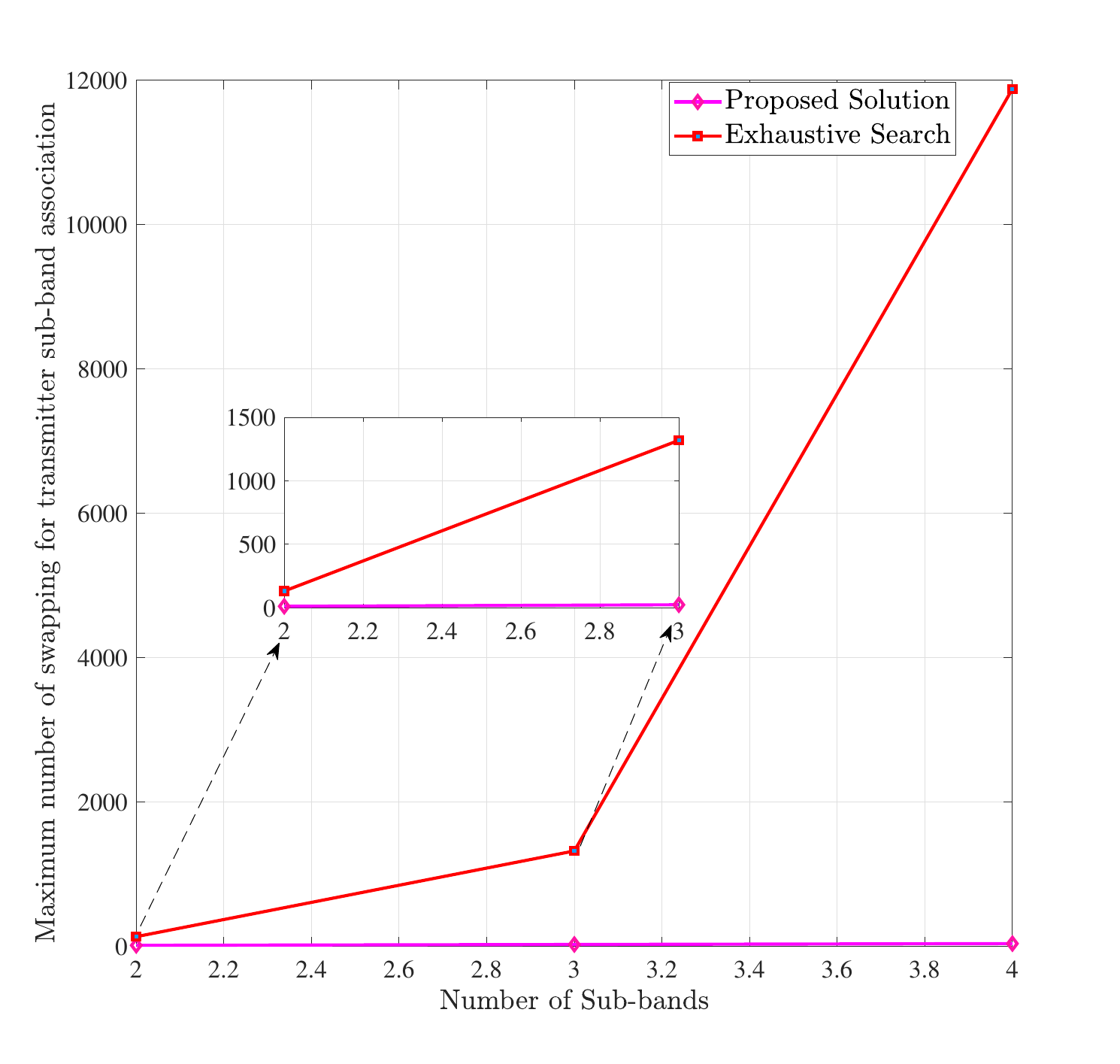}
\caption{Number of swapping required in swap matching algorithm as a function of different number of SBs.}
\label{Number_Iteration_CDF}
\end{figure}

Finally, Fig. \ref{Number_Iteration_CDF} illustrates the number of iterations required for the swap matching to reach convergence. The diagram supports the assertion made in Section V's stability analysis that there is a maximum number of iterations beyond which further enhancements in system efficiency are not possible. 
The figure illustrates that as the number of SBs grows, the maximum number of iterations required for the algorithm to converge also increases, which is reasonable. This is because a large number of SBs enables more TXs to function concurrently, as each SB permits one TX and its associated pair of RXs, thus accommodating more number of TXs and RXs simultaneously in the system. The increased number of RXs expands the search space for swap matching. As a result, the algorithmic convergence requires an increasing number of iterations to complete, which aligns with the findings shown in Fig. \ref{Number_Iteration_CDF}. In addition, this figure demonstrates the substantial reduction in time complexity experienced by our proposed solution compared to the exhaustive approach, particularly when dealing with more number of SBs. This can be explained as follows:
our method allocates only one node per SB, resulting in $K$ possible ways to select a node for one SB associated system configuration, where $K$  is the total number of system nodes. As the number of SBs grows, so does the number of selected nodes. Assuming the SB count increases from 1 to $N$, and maintaining the rule of one node per SB, the first SB can be filled in $K$ ways. The second SB then has $(K-1)$ options for node selection, and this pattern continues for $N$ number of SBs. Consequently, the total number of possible node combinations for $N$ SBs is expressed as $(K\times(K-1)\times(K-2)\times......\times(K-N+1))$, which equates to ${K\choose N}\times K!$. The trend observed in Fig. \ref{Number_Iteration_CDF} can be justified by examining how the time complexity changes as $N$ varies from two to four.

\section{Stability Analysis}
The stability of the proposed swap matching algorithm is a critical factor in ensuring its effectiveness and reliability for resource allocation and delay minimization. This section examines the stability properties of the algorithm, asserting that it achieves a stable state after a finite number of swap operations, beyond which no further improvements in utility can be made.

To substantiate this claim, consider the progression of the swap matching solution through a sequence of configurations, denoted as \( \Psi_1 \rightarrow \Psi_2 \rightarrow \Psi_3 \rightarrow \dots \). At each stage of this sequence, the system’s utility—quantified in terms of overall transmission delay—either improves or remains constant. This behavior is rooted in the Pareto-optimality preservation principle embedded within the swapping algorithm. According to this principle, a swap is executed only if it benefits at least one participant without reducing the utility of others. Consequently, the utility at a subsequent state always satisfies \( U_{\Psi_{i+1}} \geq U_{\Psi_{i}} \), ensuring a non-decreasing utility pattern throughout the process.

More generally, the improvement in utility between successive phases, denoted as \( \mathbb{U}_{l \rightarrow (l+1)} = U_{\Psi_{l+1}} - U_{\Psi_{l}} \), is non-negative (\( \mathbb{U}_{l \rightarrow (l+1)} \geq 0 \)). This property guarantees that the algorithm progressively approaches an optimal configuration, with each swap contributing to either an improvement or maintenance of the current utility. Given the finite number of nodes and SBs in the system, there is an inherent upper bound on the number of feasible swaps. Specifically, the finite number of nodes implies a limited set of possible configurations for matching nodes to SBs. 
These constraints ensure that the algorithm cannot perpetually perform swaps and will eventually reach a terminal state where no further beneficial swaps are possible.

At this terminal state, the system achieves its minimum possible information exchange delay, signifying the stabilization of the access delay. The stability of the algorithm is thus directly tied to the convergence of the swap matching process. The stabilization criterion can be expressed as follows: after a finite sequence of swaps, the utility \( U_{\Psi} \) reaches a maximum value beyond which no subsequent swap can yield a further improvement. This ensures that the algorithm halts in a stable and optimal configuration. The stable convergence of the algorithm is a significant advantage, as it guarantees that the proposed swap matching approach is both computationally efficient and practically implementable. By ensuring that the utility is always non-decreasing and bounded by the finite configuration space, the algorithm provides a robust and scalable solution for minimizing information exchange delays in NOMA-assisted ultra-dense networks.

 \section{Computational Complexity}

In this section, we analyze the computational complexity of the solution we have proposed. Our approach is divided into two main stages: the algorithmic operational phase and the data transmission phase. Our analysis begins with a discussion of the data transfer phase and concludes with an evaluation of the overall algorithmic complexity. The NOMA-supported information exchange protocol initiates with an arbitrary allocation of SBSs and resources. To establish the initial setup of node SB association, the system employs a stable marriage algorithm, which operates with a set time complexity, denoted as $\mathcal{O}(K^2)$, where $K$ is the number of SBSs in the system. Subsequently, a process of swap-matching takes place to minimize the task completion time. This phase consists of multiple iterative rounds to achieve the best possible arrangement between TXs and SBs. Specifically, each iteration evaluates the benefit of exchanging two SBSs and their corresponding SBs. The swap is executed only when an improvement in utility is detected compared to the previous arrangement. Therefore, both the number of iterations and the quantity of swap-matching attempts per iteration impact the complexity of the swap-matching phase \cite{Peer_Effect_Imp}. 

When the number of RXs does not exceed twice the number of TXs, represented as \( 2K \), the NOMA protocol enables each RX to establish a connection with its corresponding TX. This is made possible by the uplink NOMA technique, which allows a single TX to simultaneously communicate with multiple (in this work, up to two) associated RXs. NOMA achieves this by superimposing the transmitted signals for multiple TXs within the same frequency resource block and employing SIC at the RX side to decode the intended signals. This capability ensures efficient utilization of resources while maintaining connectivity between TXs and RXs.

In the swap-matching framework, if a previous allocation has RX \( i \) assigned to SB \( N_1 \), and RX \( j \) is currently occupying SB \( N_2 \), RX \( i \) may propose to exchange its position with RX \( j \). This proposed exchange, denoted as \( \psi_{i, N_1}^{j, N_2} \), is subject to   \emph{Improved Utility}: The swap must result in a measurable improvement in the system’s overall utility. For example, the new allocation must reduce the total transmission delay. 
This ensures that every swap contributes positively to the system’s optimization objectives. 

Given these constraints, the potential swap options for each RX are inherently limited. An individual RX \( i \), currently associated with SB \( N_1 \), has at most \( N-1 \) potential relocation options. This limitation arises because there are \( N \) total SBs, and each RX can only occupy one SB at a time. This bounded search space simplifies the evaluation process, as the algorithm does not need to consider every possible configuration but focuses only on feasible and utility-enhancing swaps. The combination of these rules—leveraging NOMA’s ability to connect TXs to multiple RXs, ensuring swaps improve utility ensures the algorithm remains computationally efficient while systematically optimizing the allocation of RXs to SBs. This structured approach allows the system to achieve a stable configuration with minimized information exchange delay, making it highly suitable for ultra-dense networks. 

\begin{table}[h]
	    \begin{center}
		\caption{Computational Complexity comparison among the schemes of user selection.}
		\label{Communication_Complexity}
		\begin{tabular}{|p{40mm}|p{37mm}|}
    	\hline
		\textbf{Time Synchronization Algorithms} &\textbf{Communication Complexity}\\
		\hline
		\hline
		Simulated annealing based matching  & $\mathcal{O}(2^{K})$ \\\hline
        Binary search based User Pairing associated solution \cite{Task_Completion_Guo} & $\mathcal{O}(K^{2.5} \log K)$ \\
        \hline
	  \textit{Complexity of our Proposed Solution}  & $\mathcal{O}(K^2)$ \\
		\hline
		\end{tabular}
		\end{center}
     \end{table}
Additionally, any given SB can accommodate a maximum of two RXs. As a result, when considering a swap matching of the configuration, $\psi_{i,N_{1}}^{j, N_{2}}$  where RX $i$ remains constant, there exist $2(K - 1)$ possible alternatives for SB-receiving node pairing. Given that there are $2K$ RXs, the maximum number of swap matchings that need to be evaluated in each iteration of the swap matching algorithm is $2K(K-1)$. This is derived from the calculation $\frac{2K \times 2 (K - 1 )}{2}$. As a result, when taking into account the total number of iterations $I_{total}$, the computational complexity of the swap matching solution outlined in Algorithm 1 (lines 8 and 9) can be represented as $(I_{total}\times2K(K-1))$. The swap matching procedure stated in Algorithm 1, has been executed $N_{max}$ times to calculate an average, with each execution incorporating a slight adjustment to the power allocation factor for strong RXs of each SBs. The parameter $\alpha_{total}$ represents the number of iterations needed to examine the smallest increment in the power allocation factor across the permissible range. As a result, the overall computational complexity of the swap matching algorithm can be represented as $(2\alpha_{total}N_{max}I_{total} (K^2-K))$. Additionally, our suggested approach in \cite{My_Paper_Time_Sync}  takes into account algorithmic operational stages, which have an overall complexity of $n_{max}(\delta)\mathcal{O}(K^2)$.  Within this equation, $K$ signifies the total number of nodes in the network, $\delta$ represents the permissible standard deviation, and $n_{max}$ indicates the maximum number of iterations allowed for the algorithm to achieve convergence, as previously discussed in our research \cite{My_Paper_Time_Sync}. Algorithm 1 explicitly indicates that the algorithmic operational phase and data transfer states operate in a sequential manner. Therefore, the overall complexity of our proposed solution outlined in Algorithm 1 is $(K^2+ 2N_{max}\alpha_{total}I_{total} (K^2-K) + n_{max}(\delta)K^2$), which can be simplified to the order of $\mathcal{O}(K^2)$. 
Unlike our recommended approach,  simulated annealing proves inefficient for large networks due to its worst-case average time complexity \cite{Matching_SA}, which grows exponentially with the number of graph nodes. This renders the algorithm impractical for solving problems involving large networks.
\section{CONCLUSION}
This work examines how message transfer delay affects our proposed low-complexity distributed timing synchronization method for dense networks. The algorithmic solution we proposed in our earlier research exhibited quicker synchronization when compared to several well-known studies in this area. Despite its very good applicability to ultra-dense networks, our former investigation overlooked the message transfer delays, which could substantially reduce the synchronization rate of our proposed algorithm. Recognizing this limitation, the current study incorporates message transfer delays to validate the applicability of the suggested synchronization framework in practical network environments. This study employs an uplink NOMA-based access protocol between the sender and its linked RXs to improve synchronization speed by decreasing the time required for information collection. The study presents a simplified approach for assigning transmitting nodes and allocating power across sub-bands in NOMA-driven systems, recognizing the considerable intricacy of resource distribution in these networks. As demonstrated in Table \ref{Communication_Complexity}, our suggested approach offers a \textit{low complexity} when compared to current established resource allocation solutions. Numerical investigations confirm the strong potential of the NOMA-based solution over the serial transmission scheme. Performed study also illustrates how network synchronization speed varies with system parameters such as network connectivity, network size, number of resource blocks available in the system, and channel conditions. Unlike traditional master/slave configuration-based synchronization techniques that rely on GPS as an external time reference source, our proposed algorithm operates independently of any external timing information. This proposed approach can act as an alternative synchronization method in future dense urban network scenarios, especially when devices are unable to sync with macro base stations or other external time sources. Furthermore, its decentralized nature makes it particularly suitable for self-organizing networks, encompassing device-to-device (D2D) communication, unmanned aerial vehicles (UAV), and vehicle-to-anything (V2X) communications.
\section{ACKNOWLEDGEMENT}
This study was partially supported by the following grants: Taighde Éireann – Research Ireland Grant 13/RC/2077\_P2, EU MSCA Project COALESCE Grant No. 101130739, and the US-Ireland R\&D Partnership Programme RI-SFI-23/US/3924.
\bibliographystyle{IEEEtran}
\bibliography{main}
\vspace{-3 cm}
\begin{IEEEbiography}
[{\includegraphics[width=1in,height=1.25in,clip,keepaspectratio]{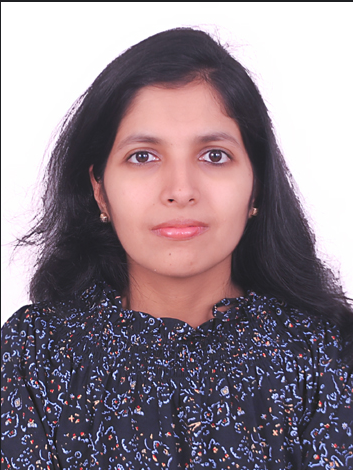}}]{Debjani Goswami}
 received the B.Tech. degree in
electronics and communication engineering from
West Bengal University of Technology, Kolkata,
India, in 2013, and the M.Tech. degree (with the
Gold Medal) in telecommunication engineering from
the National Institute of Technology, Durgapur,
India, in 2015, and the Ph.D. degree in wireless communication networks with the G. S. Sanyal School of
Telecommunications, Indian Institute of Technology
Kharagpur, Kharagpur, India, in 2022.
She is currently an Assistant Professor with the
National Institute of Technology, Calicut. Her research interests
include performance evaluation and radio resource management of beyond 5G networks, and self-organization solutions for communication networks.
\end{IEEEbiography}
\vspace{-3 cm}
\begin{IEEEbiography}
[{\includegraphics[width=1in,height=1.25in,clip,keepaspectratio]{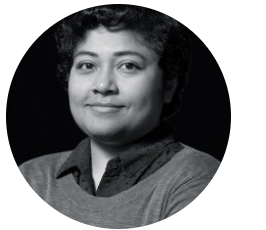}}]{Indrakshi Dey}
received the M.Sc. degree in
wireless communications from the University of Southampton,
Southampton, UK, in 2010, and the Ph.D. degree in electrical
engineering from the University of Calgary, Calgary, Canada,
in 2015. She is currently the Head of Division of the Programmable Autonomous Systems (PAS) research unit at the Walton
Institute of Information and Communications Science, Waterford, Ireland. She is currently also an Adjunct Assistant Professor
with the School of Engineering, Trinity College Dublin, Ireland
and a Science Foundation of Ireland Funded Investigator. Her
research spans over developing propagation models, analysing
performance bounds and designing communication techniques
for a variety of environments from half-duplex wireless networks,
wireless sensor networks, full-duplex networks, IoT to underwater acoustics, space and quantum communication networks.
\end{IEEEbiography}
\vspace{-3 cm}
\begin{IEEEbiography}
[{\includegraphics[width=1in,height=1.25in,clip,keepaspectratio]{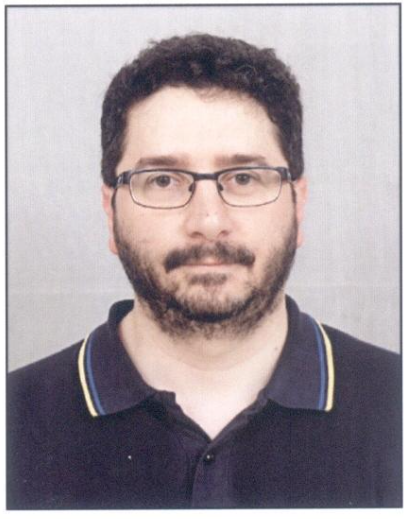}}]{Nicola Marchetti}
is Professor in Wireless Communications at Trinity College Dublin, Ireland. He an IEEE Senior Member, a Fellow of Trinity College, and was an IEEE Communications Society Distinguished Lecturer. He received the PhD in Wireless Communications from Aalborg University, Denmark in 2007, the MSc in Electronic Engineering from University of Ferrara, Italy in 2003. He has authored more than 190 journals and conference papers, 2 books and 9 book chapters, holds 4 patents, and received 4 best paper awards. His research interests span Complex Networks, Mathematics for Communications \& Computing, Network Resource Allocation. He serves as Technical Editor for IEEE Wireless Communications, and has served as an Associate Editor for IEEE Network, the IEEE Internet of Things Journal and the EURASIP Journal on Wireless Communications and Networking. 
\end{IEEEbiography}
\vspace{-3 cm}
\begin{IEEEbiography}
[{\includegraphics[width=1in,height=1.25in,clip,keepaspectratio]{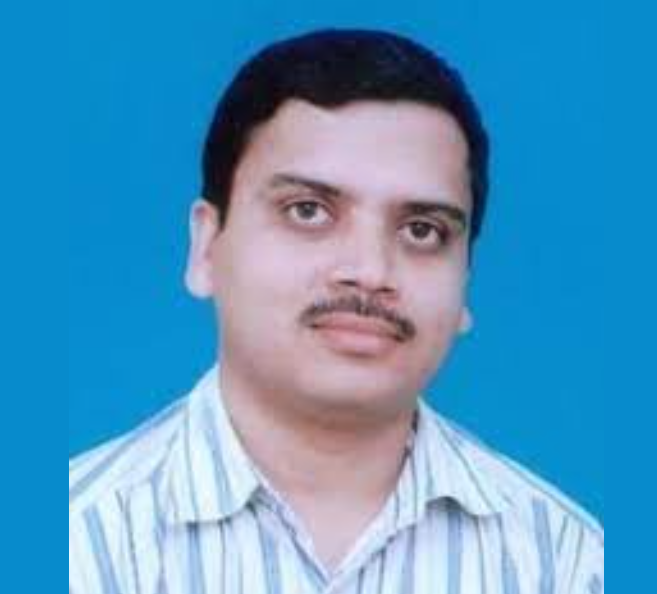}}]{Suvra Sekhar Das}
received the B.Eng. degree
in electronics and communication engineering from
the Birla Institute of Technology, Ranchi, India, in
2000, and the Ph.D. degree in wireless communication from Aalborg University, Aalborg, Denmark, in 2007. He was a Senior Scientist with the Innovation
Laboratory, Tata Consultancy Services, Kolkata,
India, from 2000 to 2008. He is currently a
Professor with the G. S. Sanyal School of
Telecommunications, Indian Institute of Technology
Kharagpur, Kharagpur, India. His research interests include the field of radio access technology.
\end{IEEEbiography}

\end{document}